\documentstyle[prl,aps,preprint,tighten,floats,epsf,rotate]{revtex}

\newbox\rotbox

\begin{document}
\draft
\preprint{
\vbox{To appear in Phys. Rev. D --Brief Reports\null\hfill TRI-PP-96-13\\
\null\hfill hep-ph/9608303}}
\title{A NOTE ON THE EXTERNAL-FIELD METHOD \\IN QCD SUM RULES}
\author{Xuemin Jin\thanks{%
Present address: Center for Theoretical Physics,
Laboratory for Nuclear Science and Department of Physics,
Massachusetts Institute of Technology, Cambridge, 
Massachusetts 02139, USA}}
\address{TRIUMF, 4004 Wesbrook Mall, Vancouver, B.C. V6T
2A3, Canada\\
and Institute for Nuclear Theory, 
University of Washington, Seattle, WA 98195, USA}
                                             
\maketitle

\begin{abstract}
The external-field method has been used extensively in the QCD
sum-rule approach to explore various hadron static properties. 
In the traditional formalism of this method, the transitions 
from the ground state hadron to excited states are not 
exponentially suppressed relative to the ground state
term and thus contaminate the ground state hadron property to be
extracted. In this paper, we suggest a modified formalism, in 
which the transition terms are exponentially suppressed relative to 
the ground state term. As such, the pole plus continuum spectral model, 
traditionally invoked in QCD sum-rule approach, can be adopted.
Thus, this modified formalism has potential to improve the 
predictability and reliability of external-field sum-rule calculations,
which is illustrated in an explicit example.
\end{abstract}
%
\narrowtext
The QCD sum-rule approach \cite{svz79} is a useful tool of extracting 
qualitative and quantitative information about hadron properties. One 
of the extensions of this approach made by Ioffe and Smilga \cite{ioffe84}
for external-field problems enables one to study hadron matrix elements 
of various currents and corresponding hadron static properties
\cite{ioffe84,ioffe83,chiu86,pasupathy87,chiu87,belyaev83,belyae85%
,chiu85,gupta89,balitsky90,henley92,ioffe92,belyaev86,jin93,jin95,jin95a,jin96}.

In the traditional external-field QCD sum-rule formalism, the 
contributions of transitions from the ground state hadron to 
excited states are not exponentially suppressed (after Borel 
transformation) relative to the ground state contribution
which contains the ground state property to be determined. 
There are, in general, infinitely many transition terms as there
are infinitely many excited states, and such terms should be retained 
in the calculations. The usual approximation adopted in the literature
is to introduce a new unknown phenomenological parameter (independent 
of Borel mass) accounting for the sum over the contributions from all 
the transitions between the ground state and the excited states. This new
parameter is extracted from the sum rules, along with the ground state
property. The isolation of the ground state property relies on that
the polynomial (in Borel mass) behavior of the ground state signal 
is different from those of the transition terms. While this approximation 
has been used in earlier studies, it has been noticed recently that the 
parameter representing the transitions, in general, should be dependent 
on the Borel mass, which can have a sizable impact on the extracted ground 
state properties \cite{ioffe95,jin95a}.

In this paper, we point out that a modified form for the linear response
of hadron correlation function to the external field can lead to  
exponential suppression of the transition terms relative to the 
ground state term. As such, the pole plus continuum spectral model,
 usually adopted in QCD sum-rule approach, can be adopted. This 
formalism has  potential to improve the predictability and reliability 
of external-field sum-rule calculations. A comparison to the traditional
formalism is illustrated in an explicit example.

The external-field method in QCD sum rules starts from a correlation 
function of hadron interpolating fields in the presence of a constant 
external field \cite{ioffe84,ioffe83,chiu86,pasupathy87,chiu87,belyaev83,belyae85%
,chiu85,gupta89,balitsky90,henley92,ioffe92,belyaev86,jin93,jin95,jin95a,jin96}
\begin{equation}
\Pi (S, q) \equiv i\int d^4 x e^{i q\cdot x}
\langle 0| T \left[ \eta_H (x) \overline{\eta}_H (0) \right] |0\rangle_S\ ,
\label{cor-gen}
\end{equation}
where $S$ denotes the external field and $\eta_H$ is a hadron 
interpolating field constructed from local QCD quark and gluon operators, 
carrying the quantum numbers of the hadron under investigation. The
presence of the external field implies that $\Pi (S, q)$ is evaluated
with an additional term
\begin{equation}
\Delta {\cal L} \equiv - S \cdot J\ ,
\label{dl}
\end{equation}
added to the usual QCD Lagrangian, where $J$ denotes the current of interest. 
Hereafter, all possible Lorentz indices of $J$ and $S$ will be suppressed.

To proceed, one usually expands Eq.~(\ref{cor-gen}) to first order in the 
external field 
\begin{equation}
\Pi (S, q) = \Pi_0 (q) + \Pi_1(q) \cdot S + \cdots \ ,
\label{expan}
\end{equation}
where $\Pi_0(q)$ is the correlation function in the absence of the external
field, which gives rise to the usual sum rules for hadron masses.
One then focuses only on the {\it linear response} term $\Pi_1 (q)$ which can 
be expressed as
\begin{equation}
\Pi_1(q) = \int d^4 x e^{i q\cdot x}
\langle 0| T \Biggl\{ \eta_H(x)\, 
\Biggl[\, \int d^4 y J(y) \,\Biggr]\,
\overline{\eta}_H(0) \Biggr\}|0\rangle \ .
\label{polar}
\end{equation}
The QCD representation of $\Pi_1(q)$, in terms of an operator product 
expansion (OPE), can be obtained by treating the external field 
perturbatively and then identifying the terms linear in $S$. 
The phenomenological representation for $\Pi_1(q)$ can be obtained by 
expanding $\Pi_1(q)$ in terms of physical hadron intermediate states 
or by invoking a dispersion relation. Here we follow the former to keep 
our discussion succinct. In a discrete-state approximation (i.e., neglecting 
the widths of hadron states, which is usually assumed in the QCD sum-rule 
approach \cite{svz79}), the phenomenological representation can be expressed
as
\begin{eqnarray}
\Pi_1(q^2) &\sim& \langle 0|\eta|h_0\rangle\,
\langle h_0| J | h_0\rangle\,
\langle h_0| \overline{\eta}_H|0\rangle\,\,
{1\over (M_0^2 - q^2)^2}
\nonumber
\\*[7.2pt]
&+&\sum_{i\neq 0}
\Biggl[\langle 0|\eta|h_0\rangle\,
\langle h_0| J | h_i\rangle\,
\langle h_i| \overline{\eta}_H|0\rangle\, +\mbox{h.c}\Biggr]\,\,
{1\over (M_0^2 - q^2) (M_i^2 - q^2)}
\nonumber
\\*[7.2pt]
&+& \mbox{terms involving only excited states}\ ,
\label{old-exp}
\end{eqnarray}
where $|h_0\rangle$ and $|h_i\rangle$ denote the ground and excited
states with mass $M_0$ and $M_i$, respectively.  
The first term contains the hadron matrix 
element, $\langle h_0| J | h_0\rangle$, of interest, the second term is 
the transition term from ground state to excited states usually discussed 
in the literature, and the rest involves only the excited states.
After a usual Borel transformation \cite{svz79}, one finds
\begin{eqnarray}
\Pi_1(M^2) &\sim&
{\lambda_0^2\over M^2}\,  \langle h_0| J | h_0\rangle\, e^{-M_0^2/M^2}
+  \sum_{i\neq 0} \Biggl[
{\lambda_0\lambda_i \langle h_0| J | h_i\rangle\over M_i^2-M_0^2}\,
\left(1-e^{-(M_i^2 - M_0^2)/M^2}\right)
+\mbox{h.c}\Biggr]\,
e^{-M_0^2/M^2}
\nonumber
\\*[7.2pt]
&+& \mbox{exponentially suppressed terms}\ ,
\label{old-form}
\end{eqnarray}
where $\lambda_0$ and $\lambda_i$ denote the coupling strengths of the 
hadron interpolating field to the ground and excited states, respectively,
and $M$ is the Borel mass.

It can be seen clearly that the second term in Eq.~(\ref{old-form}) is not 
exponentially damped as compared to the first term. This has been stressed
constantly in the literature \cite{ioffe84,ioffe83,chiu86,pasupathy87,chiu87,%
belyaev83,belyae85,chiu85,gupta89,balitsky90,henley92,ioffe92,belyaev86,jin93,%
jin95,jin95a,jin96}. The usual strategy is to approximate the second (transition) 
term as $A e^{-M_0^2/M^2}$ with $A$ a constant to be extracted from the sum rule
along with the hadron matrix element of interest. However, one notices that 
$A$, in general, is a complicated function of the Borel mass 
$M^2$. The neglect of such a Borel mass dependence will generate errors in the 
extracted hadron matrix element (see for example Refs.~\cite{ioffe95,jin95a}). 

To improve this situation, we suggest to use the combination 
$(M_0^2 - q^2) \,\, \Pi_1(q^2)$, instead of $\Pi_1(q^2)$ alone. 
The corresponding sum rule for this combination has the form
\begin{eqnarray}
\widetilde{\Pi}_1(M^2) &\sim &
\lambda_0^2\,  \langle h_0| J | h_0\rangle\, e^{-M_0^2/M^2}\,\,\,
+ \sum_{i\neq 0} \Biggl[\lambda_0\lambda_i 
\langle h_0| J | h_i\rangle\,
+\mbox{h.c}\Biggr]\,
e^{-M_i^2/M^2}
\nonumber
\\*[7.2pt]
&+& \mbox{exponentially suppressed terms}
\label{new-form}
\\*[7.2pt]
&\equiv& \lambda_0^2 \,  \langle h_0| J | h_0\rangle\, e^{-M_0^2/M^2}
+ \sum_{i\neq 0} A_i\, e^{-M_i^2/M^2}\ .
\label{new-formf}
\end{eqnarray}
The transitions from ground state to excited states
(second term) are now exponentially suppressed with respect to the 
ground state (first) term. This indicates that one 
may implement the traditional pole plus continuum model for the 
spectral ansatz of $\widetilde{\Pi}_1$. That is, one may model
$\sum_{i\neq 0} A_i\, e^{-M_i^2/M^2}$ in terms of the perturbative
evaluation of $\Pi_1$, starting from an effective threshold. As such,
one no longer needs to introduce any phenomenological parameter to
represent the transitions from the ground state to excited states as their
contributions have been included in the continuum model. Therefore, this new
formalism has potential to improve the predictability and reliability
of the external-field method.

As an example of illustration, we consider the proton matrix element
of isovector-scalar current, $H \equiv \langle p|\overline{u} u
-\overline{d}d |p\rangle/2 M_p$. This matrix element has been 
studied in Ref.~\cite{jin95} within the traditional external-field
QCD sum-rule method (see also Ref.~\cite{jin95a}). Here, our purpose 
is only to compare the new formalism suggested above with the traditional 
formalism. Thus, we shall adopt the same proton interpolating field and 
notations as used in Ref.~\cite{jin95}. 

In the traditional formalism, the sum rule used to extract $H$ is 
given by \cite{jin95}
\begin{eqnarray}
& &2\,H\,\widetilde{\lambda}_p^2\,{M_p\over M^2}\,e^{-M_p^2/M^2}
+\widetilde{A}\, e^{-M_p^2/M^2} 
\nonumber
\\*[7.2pt]
& &\hspace*{1cm} = - 2 a M^2 E_0 L^{-4/9}\, + {8\over 3} \chi a^2 L^{4/9}\,
+{1\over 3} m_0^2 a L^{-8/9}\, - { 2 \chi\over 3 M^2} a^2 m_0^2 L^{-2/27}\ ,
\label{sr-old}
\end{eqnarray}
where $M_p$ is the proton mass,
$L = \ln(M^2/\Lambda_{\rm QCD}^2)/\ln(\mu^2/\Lambda_{\rm QCD}^2)$,
and $E_0 = 1 - e^{-s_0/M^2}$ with $s_0$ the continuum threshold. 
The values of $a$, $m_0^2$, $\Lambda_{\rm QCD}$,
and $\mu$ can be found in Ref.~\cite{jin95}. In the new formalism of 
Eq.~(\ref{new-formf}), the sum rule becomes 
\begin{eqnarray}
2\,H\,\widetilde{\lambda}_p^2 \, M_p\,e^{-M_p^2/M^2}
&=& -2\, a \, (M_p^2 E_0 - M^2 E_1)\, M^2 L^{-4/9}\, 
+ {8\over 3} \chi\, a^2\, M_p^2 L^{4/9} 
\nonumber
\\*[7.2pt]
& & +{1\over 3} m_0^2\, a\, M_p^2 \, L^{-8/9}
- { 2 \chi\over 3} \left(1+{M_p^2\over M^2}\right)\, a^2\, m_0^2 L^{-2/27}\ ,
\label{sr-new}
\end{eqnarray}
where $E_1 = 1 - (1+s_0/M^2) e^{-s_0/M^2}$.
In Eq.~(\ref{sr-old}), the transition term from the proton to 
excited states is not exponentially suppressed relative to the 
ground state term and is approximated as a constant phenomenological 
parameter $\widetilde{A}$ multiplied by $e^{-M_p^2/M^2}$. This parameter 
is to be determined from the sum rule along with the matrix element $H$ of 
interest. On the other hand, there is no such parameter in Eq.~(\ref{sr-new}) 
as the transition term is exponentially suppressed with respect to the ground 
state term [see Eq.~(\ref{new-formf})] and its contribution is included 
in the continuum model. 

To extract $H$ from the above sum rules, we follow the numerical optimization 
procedure adopted in Refs.~\cite{jin95,jin95a}. For definiteness, we take 
for $\chi$ the value of $\chi = 2.2 $ GeV$^{-1}$ and the same Borel window 
as used in Ref.~\cite{jin95}. In the traditional method, the continuum threshold 
$s_0$ is taken to be the same as that in the mass sum rules and $H$
and $\widetilde{A}$ are extracted from the sum rule
\cite{ioffe84,ioffe83,chiu86,pasupathy87,chiu87,belyaev83,belyae85%
,chiu85,gupta89,balitsky90,henley92,ioffe92,belyaev86,jin93,jin95}. 
We use $\widetilde{\lambda}_p^2 \simeq 1.85$ GeV$^6$ and 
$s_0 \simeq 2.14$ GeV$^2$ \cite{jin95a}, which are slightly different 
from those used in Ref.~\cite{jin95}.
In the new formalism, however, we extract both $H$ and the continuum 
threshold $s_0$ from the sum rule as the contribution of the transitions 
is effectively included in the continuum model. The predictions turn out to be
\begin{eqnarray}
H &\simeq& 0.60\hspace*{1cm} [\mbox{from Eq.~(\ref{sr-old})}]\ ,
\label{res-old}
\\*[7.2pt]
H &\simeq& 0.97\hspace*{1cm} [\mbox{from Eq.~(\ref{sr-new})}]\ .
\label{res-new}
\end{eqnarray}
So, the two sum rules give very different predictions for the 
matrix element $H$. Which one is more reliable ? 

We have demonstrated in Eq.~(\ref{old-form}) that the 
parameter $\widetilde{A}$ in the traditional formalism should be 
a complicated function of the Borel mass. Ignoring this Borel
mass dependence will alter the curvature of the phenomenological 
side of the sum rule and hence generate errors in sum-rule 
predictions. This, for the example under consideration,
has been discussed in Ref.~\cite{jin95a} (see also Ref.~\cite{ioffe95}). 
In particular, a {\it complete} form for the phenomenological side,
which must be invoked in the traditional formalism, has been derived,
and the corresponding sum rule takes the form~\cite{jin95a}
\begin{eqnarray}
& &2\,H\,\widetilde{\lambda}_p^2\,{M_p\over M^2}\,e^{-M_p^2/M^2}
+\widetilde{A}\, e^{-M_p^2/M^2} 
+ \widetilde{B}\, {s_0^2\over 2}\,L^{-8/9}\, e^{-s_0/M^2}
\nonumber
\\*[7.2pt]
& &\hspace*{1cm} = - 2 a M^2 E_0 L^{-4/9}\, + {8\over 3} \chi a^2 L^{4/9}\,
+{1\over 3} m_0^2 a L^{-8/9}\, - { 2 \chi\over 3 M^2} a^2 m_0^2 L^{-2/27}\ .
\label{sr-3}
\end{eqnarray}
It is the last term on the left-hand side which is ignored in the traditional
formalism Eq.~(\ref{sr-old}). This term is Borel mass dependent and has
impact on the prediction for $H$. There are three parameters $H$, 
$\widetilde{A}$, and $\widetilde{B}$ to be extracted from this sum rule,
and $s_0$ is to be fixed at its value in the mass sum rules. 
Using the same numerical procedure, we find
\begin{equation}
H \simeq 0.98\hspace*{1cm} [\mbox{from Eq.~(\ref{sr-3})}]\ ,
\label{res-3}
\end{equation}
which is very close to the result from the new formalism [Eq.~(\ref{res-new})].
This indicates that the traditional treatment of the transition term has
serious drawbacks arising from the approximation adopted, which can lead to
large errors in the sum-rule predictions for the ground state matrix element. 
The new formalism, on the other hand, overcomes these drawbacks and greatly 
improves the reliability of the sum-rule predictions. 

\begin{figure}[t]
\begin{center}
\epsfysize=11.7truecm
\leavevmode
\setbox\rotbox=\vbox{\epsfbox{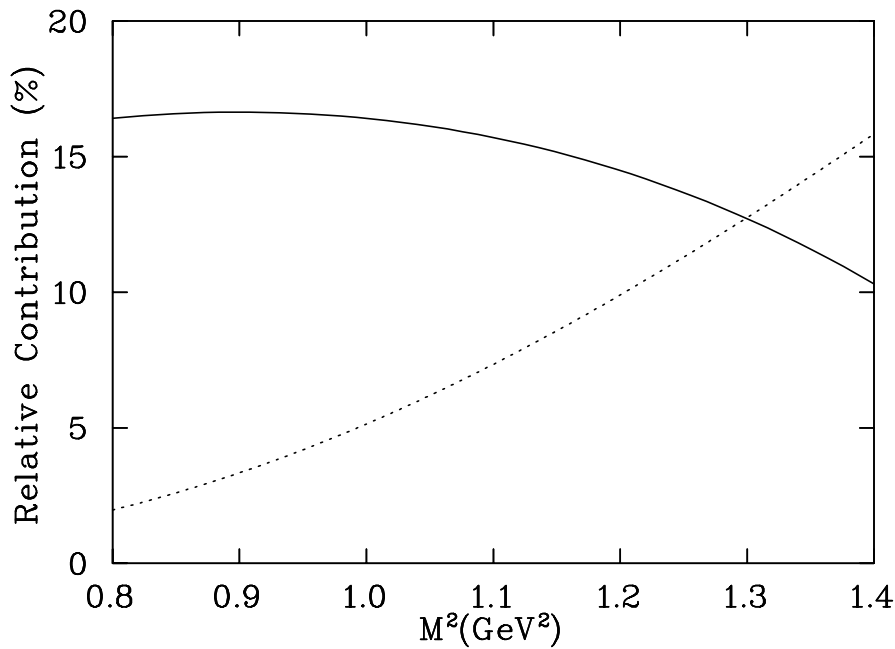}}\rotl\rotbox
\end{center}
\caption{The relative contribution (solid curve) of the transition term
plus continuum to the total phenomenological side in the traditional 
formalism Eq.~(\protect{\ref{sr-old}}) compared to that (dotted curve)
of the continuum to the total phenomenological side in the new formalism
Eq.~(\protect{\ref{sr-new}}).}
\label{fig-1}
\end{figure}

One notices that the contribution of continuum in Eq.~(\ref{sr-new}) differs 
from that in Eq.~(\ref{sr-old}). This rises the question as to whether the 
change in the continuum contribution will affect the accuracy of the new 
formalism. We note that in the new formalism the contributions from the 
transitions are included in the continuum model. It is thus fair to make a 
comparison between the relative contribution of the continuum to the total 
phenomenological side in the new formalism and the relative contribution 
of the transition term plus continuum to the total phenomenological
side in the traditional formalism. The result is shown in Fig.~\ref{fig-1}.
We see that the contribution of the continuum in the new formalism is smaller
than that of the continuum plus transitions in the traditional formalism
in most of the Borel window. This can also be seen from that a term of 
the continuum contribution in the traditional formalism becomes two terms 
in the new formalism which tend to cancel each other[see Eq.~(\ref{sr-old}) 
and (\ref{sr-new})]. Thus, we conclude that the change in the continuum contribution
does not affect the accuracy of the sum-rule prediction in the new formalism. 

Since the result from the new formalism is very similar to that from the
traditional formalism {\it with} a complete form for the phenomenological
representation [Eq.~(\ref{sr-3})], one may wonder the advantages of the 
new formalism. An obvious one is that the usual pole plus continuum
model can be used in the new formalism, which reduces
the number of phenomenological parameters to be extracted from the sum rules
[e.g., $H$ and $s_0$ in the new formalism vs. $H$, $\widetilde{A}$, and 
$\widetilde{B}$ in Eq.~(\ref{sr-3})]. A more important advantage is that
the new formalism can be applied to external-field problems with any
kind of external fields (vector, tensor, etc.) while a complete form 
for the phenomenological representation needed in obtaining Eq.~(\ref{sr-3})
cannot be derived explicitly in the case of non-scalar external fields.

The external-field method used in QCD sum-rule calculations 
concerns only the {\it linear response} term $\Pi_1(q)$ of the correlation 
function, which is actually a three-point function [Eq.~(\ref{polar})].  
So one may evaluate $\Pi_1(q)$ by following the usual external-field method
or by applying the OPE directly to the time-ordered operator product
in $\Pi_1(q)$. The latter was initiated by Balitsky and Yung \cite{balitsky83}
and followed by others \cite{braun87,he95}. The resulting sum rules in these
two apparently different approaches must be identical. The apparent differences
are only in the organization of the terms on the QCD sides of the sum rules.
However, the calculations in the external-field method are usually simpler
and more transparent than in the direct three-point function
calculations.

It is interesting to note that Balitsky and Yung \cite{balitsky83}
have proposed a procedure similar to the one discussed
in the present paper. Such procedure has been adopted in later QCD sum-rule 
calculations based directly on three-point functions \cite{braun87,he95},
but not in the external-field method. However, the authors of 
Refs.~\cite{balitsky83,braun87} stated that the transitions
from the ground state to the excited states are eliminated or canceled out
in the sum rule for $\widetilde{\Pi}_1$. This statement is obviously not true. 
The transitions from the ground state to the excited states do contribute to 
the sum rule, and their contributions are included in the continuum model. 
We also note that in Ref.~\cite{balitsky83,braun87,he95} the continuum 
threshold in the sum rule for $\widetilde{\Pi}_1(q)$ has been assumed 
to be the same as the one in the mass sum rules. This assumption, as
shown above, is unjustified. Using the same threshold in these 
two cases will introduce artificial bias to the extracted hadron properties.

As emphasized in Ref.~\cite{jin95a}, it is the product of $\lambda_H^2$
and the hadron matrix element of interest that usually appears in the 
sum rules for $\widetilde{\Pi}_1(q)$. One then needs a good knowledge of 
$\lambda_H^2$ in order to extract the interested hadron matrix elements 
cleanly. This means that uncertainties associated with $\lambda_H^2$ may 
give rise to additional uncertainties in the determination of the hadron 
matrix elements. This is a general drawback of the sum rules for $\Pi_1(q)$. 
This drawback can only be sidestepped if the external field is treated 
non-perturbatively. The reader is referred to Ref.~\cite{burkardt96} 
for more discussions on this point.

In summary, we have pointed out that a modified form for the
linear response of hadron correlation function to the external field,
concerned in the external-field method in QCD sum rules,  can lead 
to exponential suppression of the transitions from ground state to 
excited states relative to the ground state term and hence reduce
the transition contamination to the ground state property. This 
approach allows one to use the traditional pole plus continuum
spectral model and thus has potential to improve the predictability 
and reliability of external-field sum-rule calculations, which has been
illustrated in an explicit example.  Our hope is that this new 
formalism of the external-field method will generate 
new interest in revealing hadron static properties via the 
external-field QCD sum-rule method.

\vspace{21pt} 
I would like to thank the Institute for Nuclear Theory at the University 
of Washington for its hospitality and the US Department of Energy for 
partial support during the completion of this work. I am indebted to 
Derek Leinweber for insightful conversations. The support from the Natural 
Sciences and Engineering Research Council of Canada is gratefully 
acknowledged.


%
\end{document}